\documentclass[aps,pra,twocolumn,floats,showpacs,superscriptaddress]{revtex4-1}
\usepackage{graphicx,epsfig}
\usepackage{times}
\usepackage{graphics,dcolumn,bm,float}
\usepackage{amssymb,amsmath,rotate,color}
\usepackage[title,titletoc,toc]{appendix}

\usepackage[pagebackref=false,colorlinks,linkcolor=blue,citecolor=blue,urlcolor=magenta]{hyperref}

\usepackage[mathlines]{lineno}
\usepackage{hyperref}

\newcommand{\be}{\begin{equation}}
\newcommand{\ee}{\end{equation}}

\newcommand{\bearr}{\begin{eqnarray}}
\newcommand{\eearr}{\end{eqnarray}}
\newcommand{\nn}{\nonumber}

\begin{document}

\title{Enhanced weak force sensing through atom-based coherent noise cancellation in a hybrid cavity optomechanical system}
\author{S.K. Singh}
\affiliation{School of Physical Sciences, Central University of Karnataka, Kalaburagi-585367, India}
\author{M. Mazaheri}
\email{mojtaba.mazaheri@hut.ac.ir}
\affiliation{Department of Basic Science$,$ Hamedan University of Technology$,$ Hamedan$,$ 65169-1-3733$,$ Iran}
\author{Jia-Xin Peng}
\email{18217696127@163.com}
\affiliation{State Key Laboratory of Precision Spectroscopy, Quantum Institute for Light and Atoms, Department of
Physics, East China Normal University, Shanghai 200062, People’s Republic of China}
\author{M. Asjad}
\affiliation{Electrical and Computer Engineering Department, Abu Dhabi University, Abu Dhabi 59911, United Arab Emirates}

\date{\today}

\begin{abstract}
\section*{Abstract}
We investigate weak force sensing based on coherent quantum noise cancellation in a nonlinear hybrid optomechanical system. The optomechanical cavity contains a moveable mechanical mirror, a fixed semitransparent mirror, an ensemble of ultracold atoms, and an optical parametric amplifier (OPA). Using the coherent quantum noise cancellation (CQNC) process, one can eliminate the back action noise at all frequencies. Also by tuning the OPA parameters, one can suppress the quantum shot-noise at lower frequencies than the resonant frequency. In the CQNC scheme, the damping rate of the mechanical oscillator matches the damping rate of the atomic ensemble, which is experimentally achievable even for a low-frequency mechanical oscillator with a high-quality factor. Elimination of the back action noise and suppression of the shot noise significantly enhance force sensing and thus overcome the standard quantum limit of weak force sensing. This hybrid scheme can play an essential role in the realization of quantum optomechanical sensors and quantum control. 

\end{abstract}

\maketitle

\narrowtext
\section{Introduction} \label{I}
Weak force sensing at the quantum limit of sensitivity has attracted a lot of attention in recent decades \cite{braginsky1980quantum,braginsky1995quantum}. This tendency leads to the development of quantum cavity optomechanics and the possibility of using quantum phenomena in macroscopic dimensions \cite{aspelmeyer2014cavity}. Optomechanical sensors have given accurate results to measure mass \cite{bin2019mass,liu2019realization}, acceleration \cite{krause2012high,qvarfort2018gravimetry}, displacement \cite{wilson2015measurement,rossi2018measurement}, and force 
\cite{caves1980measurement,schreppler2014optically,ghobadi2015weak}. However, the measurement accuracy of optomechanical sensors is limited by the shot noise and the back-action noise, in which their competition are collectively called the standard quantum limit (SQL) \cite{bowen2015quantum}. The quantum non-demolition measurement \cite{thorne1978quantum} via quantum entanglement \cite{ma2017proposal} or squeezing \cite{clark2016observation,otterpohl2019squeezed} can be used to surpass the SQL. Also, the squeezed states are traditionally  employed to detect gravitational waves via supersensitive interferometry \cite{hollenhorst1979quantum,aasi2013enhanced,dodonov1980nondemolition}.

In the optomechanical sensor, as the intensity of the optical field increases, the measurement strength and the shot noise decrease while the back-action noise increases. Therefore, to increase the force sensitivity, it is necessary to reduce or eliminate the back-action noise. It is suggested \cite{hertzberg2010back} that measuring only a single quadrature of motion can increase the force sensitivity. In this method, the effect of the back-action noise on the orthogonal quadrature in the measurement is isolated. Woolly and Clerk \cite{woolley2013two} proposed a three-mode optomechanical system to evade the back-action noise, in which two mechanical oscillators are coupled to a single cavity mode.
Hammerer $\it{et~al.}$ \cite{hammerer2009establishing} introduced an anti-noise path in the system to cancel the back-action noise of light via quantum interference. They showed that an ancillary atomic oscillator could exhibit an opposite and equal response to the optical field in the optomechanical cavity. In this way, the CQNC process can delete the back-action noise, which is induced by the radiation pressure at all frequencies, and surpass the SQL \cite{tsang2010coherent,tsang2012evading}. The atomic ensemble can be coupled to the optomechanical systems, internally or externally, to improve the optomechanical cooling \cite{genes2009micromechanical,hammerer2010optical,genes2011atom,camerer2011realization} and create the best regime relevant to force sensing at the low mechanical frequency with a high-quality factor \cite{wimmer2014coherent}.

In addition, the interaction between the atomic ensemble and mechanical oscillator has been used to provide entanglement, entangled Einstein-Podolsky-Rosen (EPR) states, and squeezed states in the hybrid optomechanical system \cite{hammerer2009establishing}. The combination of the optomechanical system, the atomic two-level system and also the OPA can be utilized as an applicable candidate for force sensing \cite{wasilewski2010quantum,muschik2011dissipatively,polzik2015trajectories}. 
Recently, the cavity optomechanical system with the OPA is used to implement ultrahigh-precision position detection \cite{peano2015intracavity}. The advantage of this scheme is that the information is imprinted on the momentum quadrature, which is not amplified. It leads to a limited suppression of signal but simultaneously reduces the noise dramatically. Then, one can obtain the SQL without losing quantum efficiency at a mechanical resonance in the squeezed resonator. On the other hand, using the OPA, the SQL can be attained in a dissipative cavity optomechanical system \cite{huang2017robust}. So far, the hybrid optomechanical system, including the OPA and the atomic ensemble, is not used to enhance force sensing. 

He \textit{et~al.} \cite{he2020normal} used the hybrid optomechanical system, containing the degenerate OPA and the atomic ensemble, to investigate the displacement spectrum of the moveable mirror and the spectrum of squeezed cavity quadrature field. 
They showed that this hybrid optomechanical system leads to a wider spacing between the two peaks in the normal mode splitting and a more significant degree of the squeezing spectrum compared with the case of the optomechanical system with just one of the atomic ensemble or the OPA. Furthermore, the peak in the squeezed spectrum diminishes at lower temperatures due to the reduction of mechanical thermal noise.
 
Under these considerations, the present paper aims to study the CQNC force sensing in the hybrid optomechanical system containing both the OPA and the atomic ensemble. The mechanical oscillator, coupled to the atomic ensemble as the oscillator with the effective negative mass, is used for force sensing. Based on the scheme of Tsang $\it{et~al.}$ \cite{tsang2010coherent,tsang2012evading}, the coupling between the optomechanical cavity and the atomic ensemble is necessary for the CQNC process. On the other hand, tuning the OPA parameters can improve the precision of force sensing and reach the sub-SQL sensitivity at small values of the optomechanical coupling without losing quantum efficiency \cite{zhao2020weak}. The rest of the present paper is organized as follows. In section \ref{II}, we present the hybrid optomechanical system and model for optomechanical force sensing and derive the equations of motion. The central conditions for ideal CQNC and shot noise reduction are given based on the physical parameters of the system.  In section \ref{III}, we discuss force sensing and the feasibility of physical parameters of the system. A summary of our main results is provided in section \ref{IV}.

\section{Hybrid optomechanical system and model} \label{II}
\begin{figure}[t]
      \includegraphics [width =1\columnwidth]{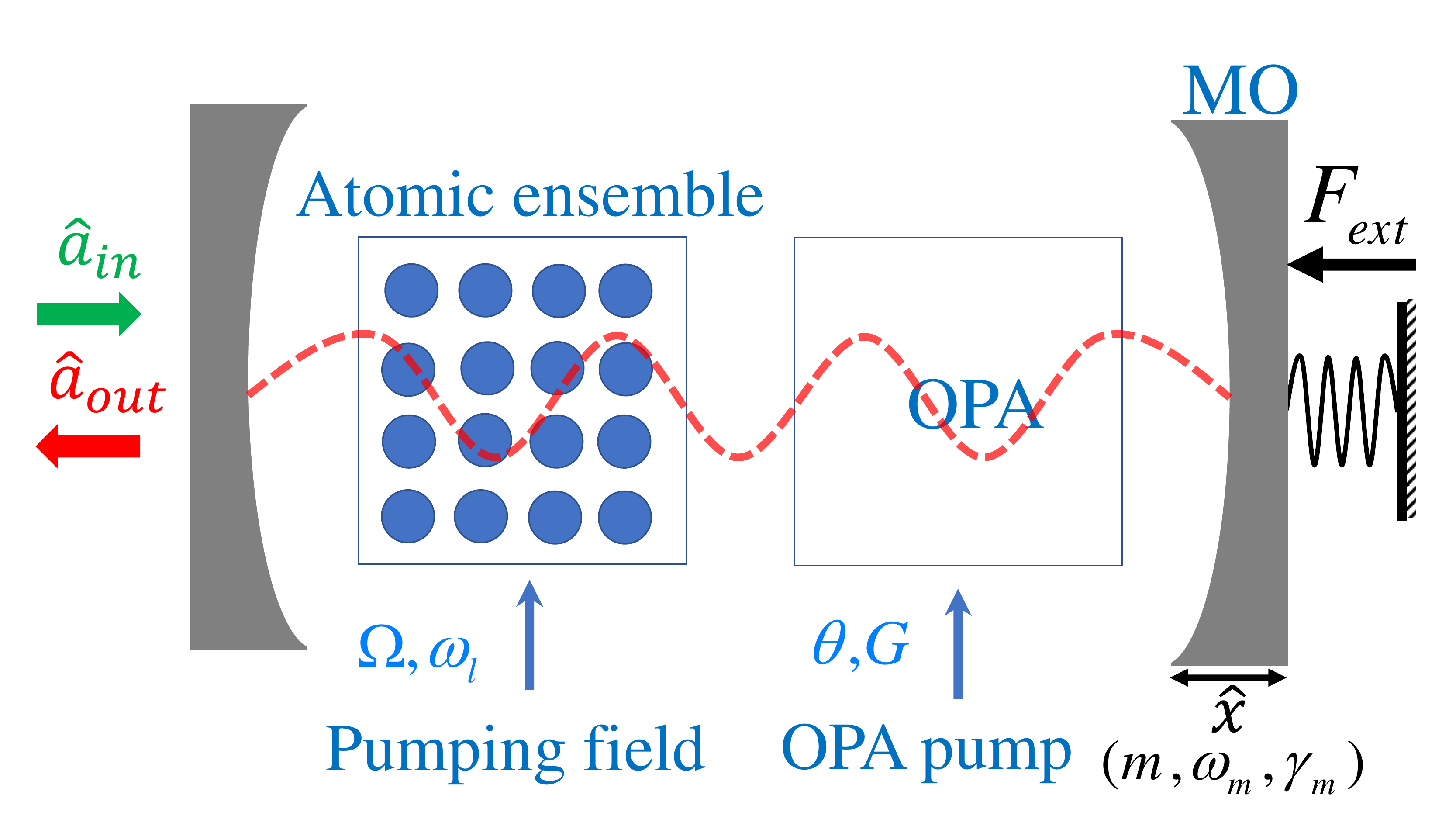}
         \caption{(Color online) Overview of the hybrid optomechanical system which consists of an optomechanical cavity, a degenerate optical parametric amplifier (OPA), and an ensemble of ultracold atoms.} \label{Fig1}
\end{figure}

We consider the hybrid optomechanical system, which contains a moveable mirror known as a mechanical oscillator, a trapped ultracold atomic ensemble and an optical parametric amplifier (OPA), as shown in Fig.~\ref{Fig1}. The atomic ensemble is driven by the intracavity field and a classical field. It can generate a strong atom-field coupling to store and transfer quantum states through the collective atomic excitation \cite{akulshin1998electromagnetically,li2004group,ian2008cavity,chang2011multistability}. The optical amplifier is used to prevent a loss in the injected squeezed light. The single-mode Fabry-Perot optical cavity includes the fixed semitransparent mirror and the movable fully reflective mirror, equivalent to a quantum harmonic oscillator with mass $m$ and frequency $\omega_m$. The optical cavity interacts with the moveable mirror through the radiation pressure. The optical cavity is driven by applying a strong laser field to the semitransparent mirror. 

Hamiltonian of the optomechanical system (OMS), Hamiltonian of the OPA \cite{zhao2020weak}, Hamiltonian of the atomic ensemble, and the Hamiltonian of the laser driving field \cite{motazedifard2019force} can be written as the following expressions: 
\begin{align}
\label{e1}
& H_{OMS}=\hbar \Delta_a a^{\dag} a+\frac{\hbar \omega_m}{2}(X^2+P^2)+\hbar g_{0} a^{\dag} a X,
\end{align} 
\begin{align}
\label{e2}
H_{OPA}=i \hbar G(e^{i\theta}a^{\dag2} - e^{-i\theta}a^{2}),
\end{align} 
\begin{align}
\label{e3}
& H_{at}=-\hbar \omega_m d^{\dag} d+\frac{\hbar G'}{2}(a+a^{\dag})(d+d^{\dag}),
\end{align}
and 
\begin{align}
\label{e4}
& H_{L}=i\hbar E_{L}(a^{\dag}-a),
\end{align}
where $a$ is the annihilation operator of the optical field, $X$ and $P$ are the dimensionless position and momentum operators of the mechanical oscillator which are normalized to zero point motion $X_{zp}=\sqrt{\hbar /(m \omega_m)}$ and zero point momentum $P_{zp}=\hbar /X_{zp}$, respectively. The parameters $g_{0}$ and $G$ are the single photon optomechanical coupling and the nonlinear gain of the degenerate OPA, and $\theta$ is the phase of the OPA pump field. $E_{L}=\sqrt{P\kappa/\omega_{L}}$ stands for the pump rate of external driving laser in which $P$, $\omega_{L}$, and $\kappa$ are the input laser driving power, the input laser frequency, and the input photon decay rate, respectively. We also used $d$ and $G'$ to denote the effective atomic bosonic annihilation operator and the effective atom-field coupling for the atomic ensemble which are thoroughly discussed in Ref. \cite{motazedifard2016force}. Hamiltonian of the system in the interaction picture containing the two level atomic ensemble and the OPA can be written as  
\begin{align}
\label{e5}
& H_{I}=\hbar \Delta_a a^{\dag} a+\frac{\hbar \omega_m}{2}(X^2+P^2)+\hbar g_{0} a^{\dag} a X \nn\\&+i \hbar G(e^{i\theta}a^{\dag2}+e^{-i\theta}a^{2})-\hbar \omega_m d^{\dag} d
\nn\\&+\frac{\hbar G'}{2}(a+a^{\dag})(d+d^{\dag})+i\hbar E_{L}(a^{\dag}-a).
\end{align}
Here, $\Delta_a=\omega_{a}-\omega_{L}$ is the detuning of the optical mode from the driving laser frequency in which $\omega_{a}$ is the frequency of the optical cavity \cite{bariani2015atom,motazedifard2016force}. Using the quantum Heisenberg-Langevin equations of motion, the equations of motion for the present system can be given as the following nonlinear equations  
\begin{align}
\label{e6}
&\dot{X}=\omega_m P,\\
\label{e7}
&\dot{P}=-\omega_m X-g_{0}a^{\dag} a-\gamma_m P+\sqrt{2\gamma_m}(\hat{F}_{th}+F_{ext}),\\
\label{e8}
&\dot{a}=-(i\Delta_a+\frac{\kappa}{2})-ig_{0}X a+2Ge^{i\theta}a^{\dag} \nn\\&-\frac{iG'}{2}(d+d')+\sqrt{\kappa}a_{in},\\
\label{e9}
&\dot{d}=i\omega_m d-\frac{G'}{2}(a+a^{\dag})-\frac{\Gamma}{2}+\sqrt{\Gamma}d_{in},
\end{align}
where $\gamma_m$, $\kappa$ and $\Gamma$ are the mechanical damping rate, the cavity photon decay rate and the collective atomic dephasing rate, respectively. The mechanical noise term is $\hat{F}_{th}$, and $F_{ext}$ is the external force to be measured which is applied to the mechanical oscillator, as shown in Fig. \ref{Fig1}. The external force is normalized to $\sqrt{\hbar m \omega_m \gamma_m}$ that is expressed in units of $\sqrt{\rm Hz}$. The equations of motion contain the dissipation and fluctuation noise terms associating with the parameters of the system. 

In the presence of strong optical drive, the quantum dynamics of the system can be well described by the linearizing the nonlinear Heisenberg-Langevin equations of motion, as shown in Refs. \cite{genes2008ground,vitali2007optomechanical}. Therefore, we expand each operator, in Eqs. \eqref{e6}-\eqref{e9}, as a steady state part and a quantum fluctuating part, as discussed theoretically \cite{huang2017robust,gebremariam2020enhancing} and experimentally \cite{safavi2013squeezed,shomroni2019two}. Then, we suppose that each operator in Eqs. \eqref{e6}-\eqref{e9} contains the steady state values ($\bar{O}$) and the fluctuating part ($\hat{O}$), including $P=\bar{P}+\hat{P}$, $X=\bar{X}+\hat{X}$, $a=\alpha+\hat{a}$, and $d=\bar{d}+\hat{d}$. We take the amplitude of intra cavity mode as a c-number, $\alpha= \vert{\alpha}\vert e^{i\phi}$, with a phase $\phi$. In order to summarize the equations of motion, the standard position and momentum quadratures of the cavity and atomic fields are defined as $\hat{x}_{i}=(\hat{a}_{i}+\hat{a}_{i}^{\dag})/\sqrt{2}$ and $\hat{p}_{i}=(\hat{a}_{i}-\hat{a}_{i}^{\dag})/(i\sqrt{2})$, and in addition, there are similar definitions for the amplitude and phase of the input quantum noises as $\hat{x}_{i}^{in}=(\hat{a}_{i}^{in}+\hat{a}_{i}^{in,\dag})/\sqrt{2}$ and $\hat{p}_{i}^{in}=(\hat{a}_{i}^{in}-\hat{a}_{i}^{in,\dag})/(i\sqrt{2})$, where $i$ stands for the cavity field mode ($\hat{a}$) and the atomic field mode ($\hat{d}$).

After straightforward calculations in the strong driving regime ($\\vert{\alpha}\vert\gg 1$), the quantum fluctuation parts of the  linearized quantum Heisenberg-Langevin equations of motion are obtained in terms of the standard quadratures as
\begin{align}
\label{e10}
&\dot{\hat{X}}=\omega_m \hat{P},\\
\label{e11}
&\dot{\hat{P}}=-\omega_m \hat{X}-\gamma_m \hat{P}-g \cos{\phi} \hat{x}_a+g \sin{\phi} \hat{p}_a \nn\\& +\sqrt{2\gamma_m} (\hat{F}_{th}+F_{ext}),\\
\label{e12}
&\dot{\hat{x}}_a=g \sin{\phi} \hat{X}+c_{-} \hat{x}_a+s_{+}\hat{p}_a+\sqrt{\kappa}\hat{x}^{in}_a,\\
\label{e13}
&\dot{\hat{p}}_a=-g \cos{\phi} \hat{X}+s_{-}\hat{x}_a-c_{+}\hat{p}_a-G'\hat{x}_d+\sqrt{\kappa}\hat{p}^{in}_a,\\
\label{e14}
&\dot{\hat{x}}_d=-\frac{\Gamma}{2}\hat{x}_d-\omega_{m}\hat{p}_d+\sqrt{\Gamma}\hat{x}^{in}_d,\\
\label{e15}
&\dot{\hat{p}}_d=-G'\hat{x}_a+\omega_{m}\hat{x}_d-\frac{\Gamma}{2}\hat{p}_d+\sqrt{\Gamma}\hat{p}^{in}_d.
\end{align} 
where parameters are defined as $g=\sqrt{2}g_0\vert{\alpha}\vert$ known as the effective linearized optomechanical coupling, $c_{\pm}=\pm \frac{\kappa}{2}+2G\cos{\theta}$ and $s_{\pm}=\pm \Delta+2G\sin{\theta}$. The laser driving power, $P$, is proportional to $g^2$ as the measurement strength. The equations of motion, Eqs. \eqref{e10}-\eqref{e15}, describe the CQNC which are similar to the equations of motion in Refs. \cite{wimmer2014coherent,bariani2015atom}. 

The atomic ensemble, as the quantum oscillator, is coupled to the optomechanical cavity but interacts with the cavity field in the exactly opposite way with an effective negative mass. Now, in order to cancel the back action noise we assume that the system is set at the resonance condition, $\Delta=0$, the $\alpha$ is real which means the intracavity phase is zero, $\phi=0$, and finally we assume the case of zero pump phase, $\theta=0$, of the optomechanical system, then Eqs. \eqref{e10}-\eqref{e15} are changed to 
\begin{align}
\label{e16}
&\dot{\hat{X}}=\omega_m \hat{P},\\
\label{e17}
&\dot{\hat{P}}=-\omega_m \hat{X}-\gamma_m \hat{P}-g \hat{x}_a+\sqrt{2\gamma_m}(\hat{F}_{th}+F_{ext}),\\
\label{e18}
&\dot{\hat{x}}_a=(-\frac{\kappa}{2}+2G)\hat{x}_a+\sqrt{\kappa}\hat{x}^{in}_a,\\
\label{e19}
&\dot{\hat{p}}_a=-g \hat{X}+(-\frac{\kappa}{2}-2G)\hat{p}_a-G'\hat{x}_d+\sqrt{\kappa}\hat{p}^{in}_a,\\
\label{e20}
&\dot{\hat{x}}_d=-\frac{\Gamma}{2}\hat{x}_d-\omega_{m}\hat{p}_d+\sqrt{\Gamma}\hat{x}^{in}_d,\\
\label{e21}
&\dot{\hat{p}}_d=-G'\hat{x}_a+\omega_{m}\hat{x}_d-\frac{\Gamma}{2}\hat{p}_d+\sqrt{\Gamma}\hat{p}^{in}_d.
\end{align} 

It can be seen from Eq. \eqref{e19} that the optical field momentum quadrature ($p_a$) depends on the position ($X$) of the mechanical oscillator and in turn the position is controlled by the momentum ($P$) of the mechanical oscillator. The external force can be detected by monitoring the optical mode. The equations of motion, Eqs. \eqref{e16}-\eqref{e21}, can be solved for the operators in the frequency domain to calculate $p_a(\omega)$ as a function of the input noises. A quantum operator in the Fourier domain can be written as 
\begin{align}
\label{e22}
\hat{O}(\omega)=\frac{1}{\sqrt{2\pi}}\int{dt\hat{O}(t)e^{-i\omega t}},
\end{align}
then using Eq. \eqref{e22}, Eqs. \eqref{e16}-\eqref{e21} can be given in the Fourier space as 
\begin{align}
\label{e23}
&\hat{X}=\chi_m(-g\hat{x}_a+\sqrt{2\gamma_m} (\hat{F}_{th}+F_{ext})),\\
\label{e24}
&\hat{P}=\frac{i\omega}{\omega_m} \hat{X},\\
\label{e25}
&\hat{x}_a=\sqrt{\kappa}\lambda_{+}\hat{x}_a^{in},\\
\label{e26}
&\hat{p}_a=g^2\chi_m\lambda_{-}\lambda_{+}\sqrt{\kappa}\hat{x}_a^{in}-g\chi_m\lambda_{-}\sqrt{2\gamma_m} (\hat{F}_{th}+F_{ext}) \nn\\& -G'\lambda_{-}\hat{x}_d+\lambda_{-}\sqrt{\kappa}\hat{p}_a^{in} ,\\
\label{e27}
&\hat{x}_d=-\omega_{m}\chi_d \hat{p}_d+\chi_d\sqrt{\Gamma}\hat{x}^{in}_d,\\
\label{e28}
&\hat{p}_d=-G'\xi \lambda_{+}\sqrt{\kappa}\hat{x}_a^{in}-\chi'_d \sqrt{\Gamma}\hat{x}_d^{in}+\xi\sqrt{\Gamma}\hat{p}^{in}_d,
\end{align}
where, the parameters are defined as 
\begin{align}
\label{e29}
& \lambda_{\pm}=(\chi_a^{-1}\mp2G)^{-1}, \\
\label{e30}
& \xi=(i\omega+\Gamma/2+\omega_m^2 \chi_d)^{-1}.
\end{align}
Also, the effective susceptibilities can be written as 
\begin{align}
\label{e33}
&\chi_m=\frac{\omega_m}{\omega_m^2-\omega^2+i\omega \gamma_m},\\
\label{e34}
&\chi_a=\frac{1}{i\omega+\kappa/2},\\
\label{e35}
&\chi_d=\frac{1}{i\omega+\Gamma/2},\\
\label{e36}
&\chi'_d=-\omega_{m}\xi \chi_d=\frac{-\omega_m}{\omega_m^2-\omega^2+i\omega \Gamma/2+\Gamma^2/4},
\end{align}
for the mechanical oscillator, the meter optical cavity, the atomic ensemble, and the last one stands for the generalized susceptibility of the atomic ensemble, respectively. 

\section{Force sensing and CQNC} \label{III}
Solving Eqs. \eqref{e23}-\eqref{e28},we can find the phase quadrature of the cavity field, $p_a$, and finally using the standard input-output relation as
\begin{align}
\label{e37}
\hat{p}_a^{out}=\sqrt{\kappa}\hat{p}_a-\hat{p}_a^{in},
\end{align}
the detected phase quadrature of the cavity field mode can be expressed in terms of the input noises as the following expression 
\begin{align}
\label{e38}
&\hat{p}_a^{out}=-g\chi_m \lambda_{-}\sqrt{2\gamma_m\kappa} (\hat{F}_{th}+F_{ext})
\nn\\& +(g^2\chi_m+G^{'2} \chi'_d)\lambda_{+}\lambda_{-}\kappa\hat{x}_a^{in}+(\lambda_{-}\kappa-1)\hat{p}_a^{in}
\nn\\& -G'\lambda_{-}\sqrt{\kappa \Gamma}[\chi_d(\omega_m\chi'_d+1)\hat{x}_d^{in}+\chi_d'\hat{p}_d^{in}].
\end{align}
The first term in Eq. \eqref{e38} represents the thermal noise and the external force where the external force is independent of the detected frequency, the second line contains the back-action noise and the shot noise of the field, and the third line describes the atomic noise. If we choose the conditions such that    
\begin{align}
\label{e39}
g^2\chi_m+G^{'2} \chi'_d=0
\end{align}
for all frequencies, then the first term in the second line of Eq. \eqref{e38}, the back-action term, will be entirely canceled in Eq. \eqref{e38} and also in the external force measurement. In the CQNC scheme, we essentially have $g=G^{'}$ and $\chi_m=-\chi'_d$ or, in other words, the contributions to the back-action noise from the mechanical oscillator and the atomic ensemble cancel each other for all frequencies. They are the noise and anti-noise contributions to the signal obtained by assuming the use of an atomic ensemble as an oscillator with the effective negative mass. In addition to the above assumptions, we assume that the mechanical resonator and the atomic ensemble have the same damping rates ($\gamma_m=\Gamma/2$) and also the mechanical oscillator has a high-quality factor such that $\Gamma\ll \omega_m$, then we can ignore the last term in the denominator of the generalized atomic susceptibility in Eq. \eqref{e36}. Therefore, the susceptibilities of mechanics and generalized atomic ensemble completely match, which leads to the perfect coherent back-action noise cancellation.

By rewriting Eq. \eqref{e38}, the relation between the external force, $F_{ext}$, and the measured phase quadrature, $\hat{p}_a^{out}$, can be given as the following expression 
 \begin{align}
\label{e40}
F_{ext}+\hat{F}_{add}=\frac{-1}{g\chi_m \lambda_{-}\sqrt{2\gamma_m\kappa}}\hat{p}_a^{out}
\end{align}
where the added force noise, $\hat{F}_{add}$, is defined as 
\begin{align}
\label{e41}
& \hat{F}_{add}=\hat{F}_{th}-\frac{\lambda_{-}\kappa-1}{g\chi_m \lambda_{-}\sqrt{2\gamma_m\kappa}}\hat{p}_a^{in}
\nn\\& +\frac{G' \chi_d' \sqrt{\kappa \Gamma}}{g\chi_m \sqrt{2\gamma_m\kappa}}[\frac{\chi_d}{\chi_d'}(\omega_m\chi'_d+1)\hat{x}_d^{in}+\hat{p}_d^{in}]. 
\end{align}
It should be noted that the back-action noise term is canceled. Simplifying Eq. \eqref{e41}, we have 
\begin{align}
\label{e42}
& \hat{F}_{add}=\hat{F}_{th}-\frac{\lambda_{-}\kappa-1}{g\chi_m \lambda_{-}\sqrt{2\gamma_m\kappa}}\hat{p}_a^{in}
\nn\\& +[-\frac{i\omega+\Gamma/2}{\omega_m}\hat{x}_d^{in}+\hat{p}_d^{in}].
\end{align}
The added noise contains the thermal Langevin force coupled to a thermal reservoir at temperature $T$ (first term), the shot noise in the phase quadrature of the optical field (second term) and the atomic noise (third term).

Using the definition of spectral density of the added noise \cite{wimmer2014coherent}, $S_{F,add}(\omega)$, we can find the sensitivity of the force measurement as  
\begin{align}
\label{e43}
S_{F,add}(\omega)\delta(\omega-\omega')=\frac{1}{2}(\left\langle \hat{F}_{add}(\omega)\hat{F}_{add}(-\omega')\right\rangle +c.c. ),
\end{align}
then, using the conditions of the CQNC and for $\omega \ll \kappa$, we find the spectral noise as 
\begin{align}
\label{e44}
& S_{F,add}(\omega)=\frac{k_BT}{\hbar\omega_m}+\frac{1}{2} \frac{1}{g^2|\chi_m(\omega)|^2 (2\gamma_m\kappa)}|\frac{\lambda_{-}(\omega)\kappa-1}{\lambda_{-}(\omega)}|^2
\nn\\& +\frac{1}{2}(\frac{\omega^2+\Gamma^2/4}{\omega_m^2}+1).
\end{align}
After some calculation, the spectral density can be more simplified as 
\begin{align}
\label{e45}
& S_{F,add}(\omega)=\frac{k_BT}{\hbar\omega_m}+\frac{1}{2}[\frac{(\kappa/2-2G)^2}{g^2|\chi_m(\omega)|^2 (2\gamma_m\kappa)}
\nn\\&(\frac{\omega^2+\Gamma^2/4}{\omega_m^2}+1)],
\end{align}
where obviously the back-action noise term, scaling to $g^2$, is deleted. This result can be compared to that of the standard optomechanical system, discussed in Refs. \cite{aspelmeyer2014cavity,wimmer2014coherent,bariani2015atom}, as the following equation
\begin{align}
\label{e46}
 S_{F}(\omega)=\frac{k_BT}{\hbar\omega_m}+\frac{1}{2} [ \frac{\kappa}{\gamma_m}\frac{1}{g^2|\chi_m|^2}\frac{1}{4}+4g^2\frac{1}{\kappa\gamma_m}]
\end{align}
which contains the shot noise and the back-action noise terms proportional to $1/g^2$ and $g^2$, respectively. The thermal Brownian noise results in a constant background noise to the force sensitivity independent of the input laser power. 
At $T=0$, we can minimize the right hand side of Eq. \eqref{e46} with respect to $g^2$ or proportionally the input laser power, $P=2\hbar \omega_L \kappa(g/g_0)^2$, gives the SQL for continuous force sensing,
\begin{align}
\label{e47}
 S_{F,SQL}=\frac{1}{\gamma_m|\chi_m|}
\end{align}
and equivalently the minimized noise spectral density can be found for the CQNC at $T=0$ as 
\begin{align}
\label{e48}
 S_{F,CQNC}=\frac{1}{2}\frac{\omega^2+\omega_m^2+\Gamma^2/4}{\omega_m^2}.
\end{align}
The noise spectral densities are normalized by the factor, $\hbar m\omega_m\gamma_m$, then they are given in units of N$^2$Hz$^{-1}$. However, in this work, we have considered the noise spectral densities as dimensionless for simplicity. 

\begin{figure}[t]
 \includegraphics [width = 1\columnwidth]{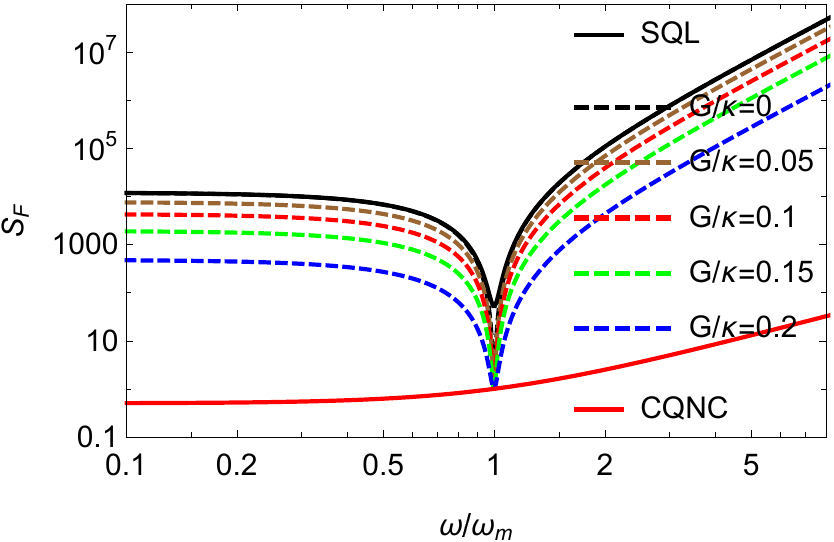}
   \caption{(Color online) Noise power spectral densities calculated for the standard optomechanical system (black solid line), $S_{SQL}$, the optomechanical system with the OPA (colored dashed line) with different $G$, and the coherent quantum noise cancellation (red dashed line), $S_{CQNC}$, in optimal power versus the detection frequency. The spectral densities are normalized to $\hbar m\omega_m\gamma_m$. The parameters are given in Table \ref{1} and we have $Q=\omega_m/\gamma_m=10^4$.} \label{Fig2}
\end{figure}

The noise spectral densities for the standard optomechanical system (SQL), the optomechanical system with OPA (dashed lines) and the CQNC schemes (CQNC) are represented in Fig. \ref{Fig2}. The spectral density for the CQNC scheme is limited to the shot noise term over the whole detection bandwidth, which is the main advantage of the CQNC scheme. Fig. \ref{Fig2} shows the variation of the noise spectral density of the optomechanical system versus the detection frequency for different values of the OPA pump gain, $G$. At the resonance condition ($\Delta=0$), the noise spectral density of the standard optomechanical system and that of the optomechanical system with the OPA equal the noise spectral density of the CQNC scheme. Fig. \ref{Fig2} shows that with the increase of relative OPA pump gain, $G/\kappa$, the SQL can be suppressed at frequencies below and as well above the mechanical resonance. The force noise spectral density is decreased by nearly two orders of magnitude on both sides of the low and high-frequency domains. Recently, Korobko \textit{et~al.} \cite{korobko2019quantum} proposed an optical approach to expand the detection bandwidth for gravitational-wave observatories in which quantum uncertainty can be squeezed inside one of the optical resonators at high frequencies. At the same time, the low-frequency sensitivity was remained unchanged. On the other hand, as mentioned in section \ref{I}, Zhao \textit{et~al.} \cite{zhao2020weak} investigated weak-force sensing in a squeezed cavity and theoretically showed that the SQL could be suppressed at frequencies below the mechanical resonance; however, the back-action noise remained unchanged in the system.  

The spectral densities are depicted in Fig. \ref{Fig3}(a) and Fig. \ref{Fig3}(b) as functions of the laser driving power, scaling to the optomechanical coupling $g^2$, for on-resonance ($\omega=\omega_m$) and off-resonance ($\omega=\omega_m+4\gamma_m$) cases, respectively. Fig. \ref{Fig3}(a) shows that at the higher driving power, the noise spectral density decreases over the bandwidth significantly. At lower values of the coupling strength, the shot noise term is dominant, and the noise spectral density of the normal optomechanical system (SQL) first decreases with the driving power increasing until the minimum point. After that, the back-action noise is dominant, and the noise spectral density of the normal optomechanical system increases. In the normal optomechanical system, the noise spectral density of the SQL represented by Eq. \eqref{e47} can be reached to the minimum point at $g_{SQL}=\sqrt{\kappa}/(2\sqrt{|\chi_m|})$, whereas this point in the normal system can be lowered by using the OPA pump with gain $G$, which gives $g_{SQ}^{\theta=0}=|\kappa-4G|/(2 \sqrt{\kappa|\chi_m|})$ \cite{zhao2020weak}. The CQNC scheme cancels the back-action noise. It improves the sensitivity in a broad bandwidth at higher values of the driving power. In the absence of OPA pumping ($G=0$), on-resonance, the CQNC system needs a higher driving power to reach the same sensitivity of the standard setup. In contrast, as the OPA pump gain $G$ increases, the sensitivity of the CQNC-OPA arrangement rapidly becomes superior to the normal optomechanical system and surpasses the SQL at lower values of the driving power. In addition to the cancellation of the back-action noise, the CQNC-OPA arrangement reduces the shot noise by about two orders of magnitude. It improves the measurement accuracy at lower driving power domain. On the other side, away from the resonance case with detuning $4\gamma_m$, the spectral density of the system, $S_{F}$, decreases with increasing the values of the OPA pump gain $G$, as depicted in Fig. \ref{Fig3}(b). We have used experimentally accessible parameters \cite{bariani2015atom,bowen2015quantum,wimmer2014coherent} in our calculations which are collected in Table \ref{1}. 

\begin{table}
\caption{Parameters of the hybrid system.}
\label{1}
\begin{tabular}{l l l}
\toprule
Parameter    &  Symbole &  Value (unit) \\
\hline
Mirror mass & $m$ & 50 ng \\
Single photon optomechanical coupling & $g_0/2\pi$ & 300 Hz \\
Mechanical resonance & $\omega_m/2\pi$ & 300 kHz \\
Mechanical damping rate & $\gamma_m/2\pi$ & 30 Hz\\
Optical cavity damping rate & $\kappa/2\pi$ & 1 MHz\\
Laser source power & $P$ & 100 mW \\
Laser frequency & $\omega_L /2\pi$ & 384 THz \\ 
\hline
\hline
\end{tabular}
\end{table}

\twocolumngrid\
\begin{center}\
\begin{figure}[h]\
\includegraphics[scale=1]{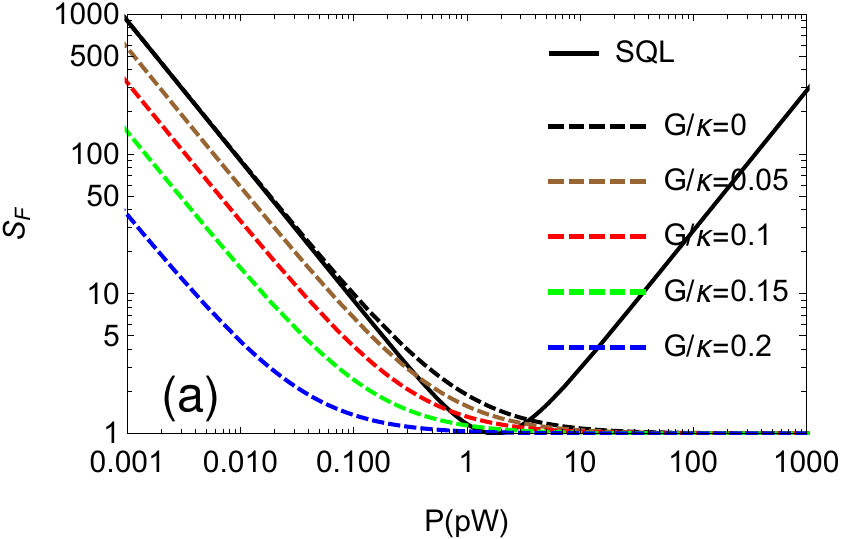}\
\includegraphics[scale=1]{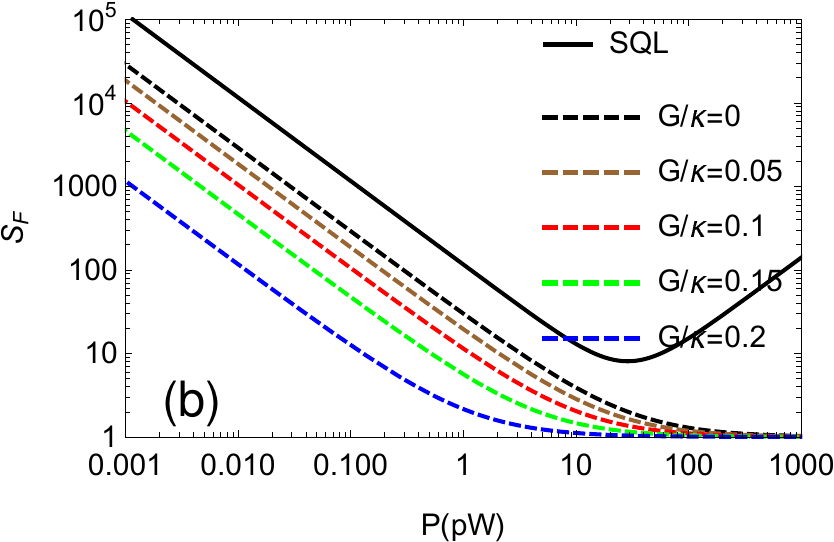}\
\caption{(Color online) Noise spectral density for weak force measurement (a) on mechanical resonance ($\omega=\omega_m$) and (b) off-resonance with $4\gamma_m$ detuning ($\omega=\omega_m+4\gamma_m$) as function of the input driving power ($P=2\hbar \omega_L \kappa(g/g_0)^2$) for the standard optomechanical system and for the present hybrid system, containing the OPA and the atomic ensemble, with different values of the relative parametric gain as $G/\kappa=0$, 0.05, 0.1, 0.15, and 0.2. The parameters are given in Table \ref{1}.}\
\label{Fig3}
\end{figure}\
\end{center}\
\twocolumngrid\  
                                                
To reach the SQL, we have considered the impact of $g^2/g^2_{SQL}$, the frequency $\omega$, and the parametric gain $G$ on the noise spectral density of the present hybrid system. Fig. \ref{Fig4}(a) and Fig. \ref{Fig4}(b) depict the noise spectral densities for the hybrid system as functions of the relative optomechanical coupling $g^2/g^2_{SQL}$ and relative frequency $\omega/\omega_m$ with $G/\kappa=0$ and 0.1, respectively. Fig. \ref{Fig4}(a) with $G/\kappa=0$ indicates that the noise spectral density is close to the minimum value in a limited range in the vicinity of the resonance, and it increases in areas farther than the resonance region. While, the noise spectral density, as shown in Fig. \ref{Fig4}(b) with $G/\kappa=0.1$, is close to the minimum value in a wider area, and it increases with a lower slope in areas farther than the resonance region, especially in areas with a frequency lower than the resonant frequency. Fig. \ref{Fig4}(a) shows that the noise spectral density reaches the minimum values in the narrow area in the vicinity of the resonant frequency, which slightly increases at higher values of $g^2/g^2_{SQL}$. While, Fig. \ref{Fig4}(b) with $G/\kappa=0.1$ shows that the spectral noise density approaches to the minimum values of the SQL in a wider range of lower values of $g^2/g^2_{SQL}$ which is interesting for feasible experimental scenarios. 

\twocolumngrid\
\begin{center}\
\begin{figure}[h]\
\includegraphics[scale=0.75]{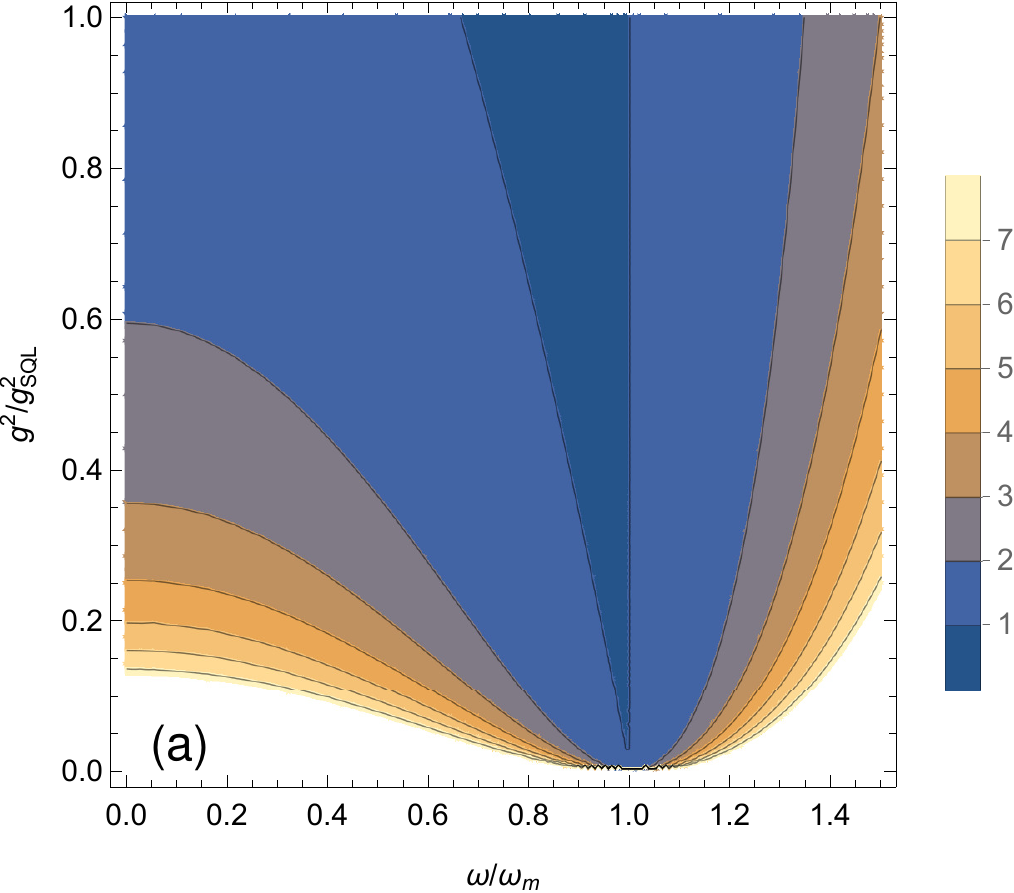}\
\includegraphics[scale=0.75]{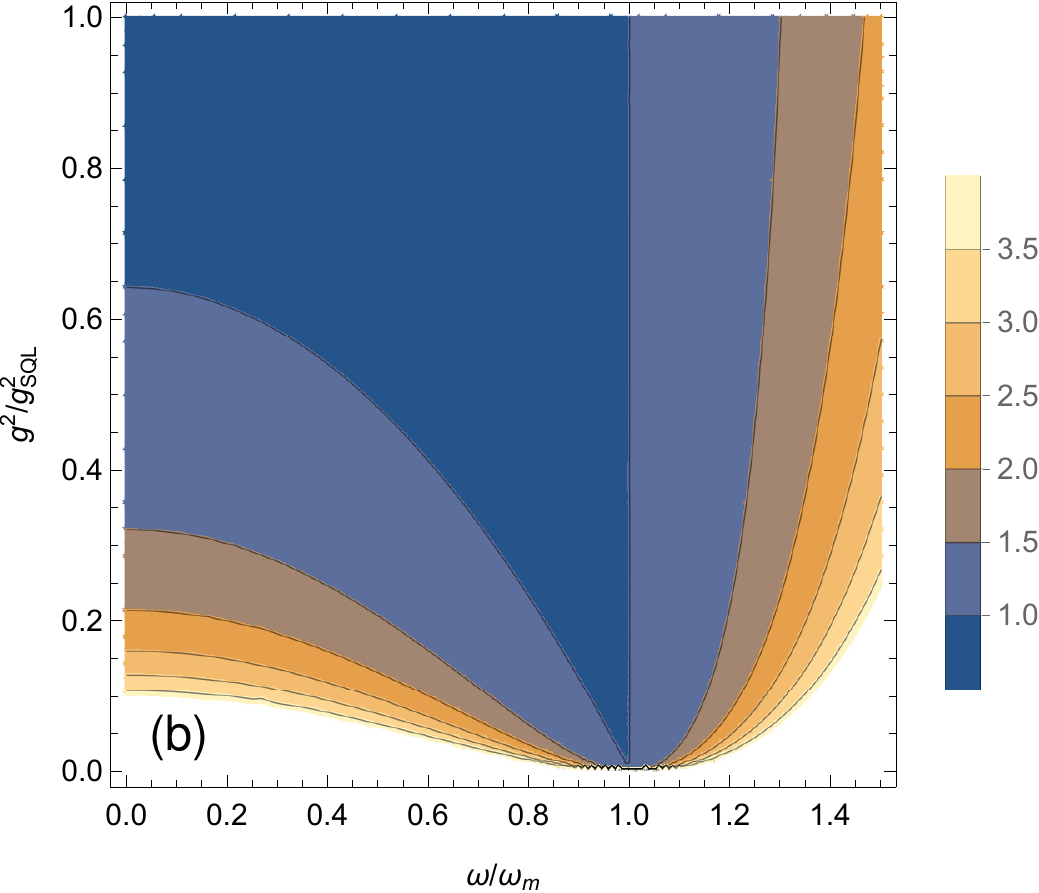}\
\caption{(Color online) Noise spectral density for weak force measurement as function of the relative squared optomechanical coupling ($(g/g_{SQL})^2$) and the relative detection frequency ($\omega/\omega_m$) for the present hybrid system, containing the OPA and the atomic ensemble with different values of the relative parametric gain as (a) $G/\kappa=0$ and (b) $G/\kappa=0.1$. The parameters are given in Table \ref{1}.}\
\label{Fig4}
\end{figure}\
\end{center}\
\twocolumngrid\

\section {Summary and conclusion} \label{IV}
In summary, we have presented the analysis of the coherent quantum noise cancellation (CQNC) and shot noise reduction scheme in the optomechanical system assisted by the ensemble of ultracold atoms and the optical parametric amplifier (OPA). The ensemble of ultracold atoms is chosen as the effective oscillator with the effective negative mass, which destructively interacts with the optical cavity mode leading to the perfect cancellation of the back-action noise. On the other hand, the quantum squeezing in the optomechanical cavity reduces shot noise in the low driving power and the broad range of detection frequency. In this hybrid system, the CQNC and shot noise reduction cooperatively occur when the effective linearized optomechanical coupling and the nonlinear gain of the OPA, the mechanical susceptibility and the generalized susceptibility of the atomic ensemble, and eventually the damping rates of the mechanical oscillator and the atomic ensemble coincide with each other, respectively.  Finally, we have also used a set of possible experimental parameters which shows the elimination of the back-action noise and the reduction of the shot noise at frequencies below the mechanical resonance. 

\section*{References}
\bibliography{paper}

\end{document}